\documentclass[12pt,a4paper, amsmath,amssymb]{article}
\usepackage[margin=0.8in]{geometry}
\usepackage{amssymb,amsmath}
\usepackage{amsfonts}
\usepackage{bm}
\usepackage[utf8]{inputenc}
\usepackage{graphicx}
\usepackage{float}
\usepackage{amsmath}
\usepackage[colorlinks,linkcolor = blue,
urlcolor  = blue,
citecolor = blue,
anchorcolor = blue]{hyperref}
\usepackage{xcolor}
\usepackage{ctable} 
\usepackage{adjustbox}
\def\thefootnote{\fnsymbol{footnote}}
\usepackage{cite}

\begin{document}
	{
\begin{center}
{\Large \textbf{Phase diagrams of  a nonlinear magnetic charged rotating AdS black hole with a quintessence field }}
\thispagestyle{empty}
\vspace{1cm}

{\sc
	
	H. Laassiri$^1$\footnote{\url{hayat.laassiri@gmail.com}},
	A. Daassou$^1$\footnote{\url{ahmed.daassou@uca.ma}},
	R. Benbrik$^1$\footnote{\url{r.benbrik@uca.ac.ma}}\\
}
\vspace{1cm}
{\sl
	$^1$Fundamental and Applied Physics Laboratory, Physics Department, Polydisciplinary Faculty,
	Cadi Ayyad University, Sidi Bouzid, B.P. 4162, Safi, Morocco.\\
}
\end{center}
\vspace*{0.1cm}
\begin{abstract}
In this paper, we investigate the phase transitions and critical behavior of a nonlinear magnetically charged rotating AdS black hole, with a particular emphasis on the influence of a quintessence field. Our comprehensive thermodynamic analysis explores the impact of thermal fluctuations on the black hole's properties. We observe that for larger black holes, the corrected entropy remains positive, similar to the uncorrected case, which highlights the significant role of thermal fluctuations in modifying the entropy of smaller black holes. Through a meticulous analysis of various thermodynamic properties, we derive explicit analytical expressions for the critical points. Our findings demonstrate that the quintessence field significantly affects phase transitions, resulting in distinct critical phenomena. Notably, the black hole's phase transitions exhibit striking similarities to those observed in van der Waals fluids, offering deeper insights into the complex thermodynamic behavior of these systems across different scales.

\end{abstract}

\def\thefootnote{\arabic{footnote}}
\setcounter{page}{0}
\setcounter{footnote}{0}

\label{intro}\
Nonlinear magnetic black holes, which come from the interaction between gravitational forces and nonlinear electrodynamics (NED), offer interesting alternatives to traditional black hole models. Unlike the linear Maxwell theory, NED allows for the creation of black holes with regular horizons, effectively avoiding the singularities typically found in standard black hole solutions \cite{1}. These nonlinear models are especially important due to their unique internal structures and stability features, making them different from those predicted solely by general relativity \cite{2}. For example, the Bardeen black hole, known as one of the first regular black hole models, can be seen as a magnetic monopole within a specific NED context, showing the potential of these solutions to explore quantum gravity effects and improve our understanding of spacetime in extreme conditions \cite{3, 4, 5}.
In Refs. \cite{6,7,8,9}, extensive discussions explored how the quintessential parameter affects the quasinormal frequencies in different black hole spacetimes. Ref. \cite{12} delved into the Bardeen solution by adding a cosmological constant and embedding it within quintessence. The study revealed that this solution can be derived by integrating nonlinear electrodynamics with the Einstein equations. The authors noted that the solution does not always maintain regularity and outlined the specific conditions required for it to be regular. They also thoroughly investigated the thermodynamic properties of this solution, including the formulation of the Smarr formula and the first law of thermodynamics in this setting. This detailed analysis provides important insights into how quintessence and nonlinear electrodynamics interact with black hole properties. Recent research on phase transitions of black holes in asymptotically anti-de Sitter (AdS) space has garnered considerable attention \cite{13,14,15}. In the context of charged AdS black holes, extensive studies have revealed the occurrence of phase transitions between small and large black hole states within the canonical ensemble, where the electric charge remains constant. This behavior mirrors the phase transitions observed in van der Waals (vdW) fluids, highlighting a striking analogy between gravitational systems and thermodynamic phenomena in classical fluids \cite{16,17,18}.

Widely used method for investigating the critical thermodynamic properties of black holes involves the analysis of phase diagrams. However, the construction of these diagrams can be particularly challenging due to the inherent complexities in certain theoretical frameworks. The black holes examined in this study present such complexities, especially when considering nonlinear magnetic charge, rotation, and the presence of quintessence along with a cosmological constant. To rigorously assess the critical behavior of this specific black hole configuration, we commence our analysis by employing the metric provided in Ref. \cite{19}
\begin{equation}\label{Eq.1}
\begin{aligned}
{\text{ds}^2=\frac{\Sigma }{\Delta _r}\text{dr}^2+\frac{\Sigma }{\Delta _{\theta }}\text{d$\theta $}^2-\frac{\Delta _r}{\Xi^2\Sigma
		}\left(\text{dt}-a \sin ^2\theta  \text{d$\phi $}\right)^2+\frac{\Delta _{\theta }\sin ^2\theta }{\Xi^2\Sigma }\left(\text{adt}-\left(r^2+a^2\right)\text{d$\phi
			$}\right)^2},
		\end{aligned}
\end{equation}
with
\begin{equation}\label{Eq.2}
\begin{aligned}
	{\Xi = 1+ \frac{\Lambda }{3}a^2, \hspace{0.5cm} \Delta _r= r^2+a^2-\frac{2\text{Mr}^4}{r^3+Q^3}-\frac{\Lambda }{3}\left(r^2+a^2\right)r^2- \text{$\beta
			$r}^{1-3\omega },\hspace{0.5cm}{S=\frac{\pi  \left(a^2+r^2\right)}{\Xi }}}.
		\end{aligned}
\end{equation}
In this context, \(Q\) and \(a\) denote the magnetic charge and rotational parameter, respectively. \(\beta\) represents the positive quintessence parameter, while the quintessence state parameter \(\omega\) is constrained within the range \(-1 < \omega < -\frac{1}{3}\) \cite{20}.

The Hawking temperature  of a black hole is related to the surface gravity by the equation \( 2\pi T = \kappa \hbar \), and is expressed as
\begin{equation}\label{Eq 3}
\begin{aligned}
T=& \left(a^2 r^{3 \omega } \left(-4 Q^3 \left(3+4 P \pi  r^2\right)+r^3 \left(-3+8 P \pi  r^2\right)\right)+3 r \left(r^{4+3 \omega
	}+8 P \pi  r^{6+3 \omega }+3 r^3 \beta  \omega +\right.\right.\\
&\left.\left.Q^3 \left(-2 r^{1+3 \omega }+3 \beta  (1+\omega )\right)\right)\right)r^{-1-3 \omega }/\left( \left(Q^3+r^3\right)12{\pi}\left(a^2+r^2\right)\right).
			\end{aligned}
\end{equation}
To evaluate the entropy corrections due to thermal fluctuations, we base our calculations on the canonical partition function.
\begin{align}\label{Eq 4}
{Z(\gamma ) = \int _0^{\infty }\sigma (E)\exp ^{-\text{$\gamma $E}}\text{dE}},
\end{align}
The symbol $\sigma(E)$ represents the quantum density of the system. By utilizing the inverse Laplace transformation and conducting detailed calculations, the density of states is derived as follows	
\begin{align}\label{Eq 13}
{\sigma (E) = \dfrac{-i e^S}{2\pi}\int _{c-\text{i$\infty $}}^{c+\text{i$\infty $}}\exp ^{\dfrac{(\gamma -c)^2}{2}\dfrac{d^2S_0}{d^2\gamma
	}}\text{d$\gamma $}}.
\end{align}
After simplifying the calculations, the corrected entropy $S_{0}$ is formulated as a function of temperature, entropy, and the corrected parameters as follows	
\begin{align}\label{Eq 15}
S_0=&{\frac{3 \pi  \left(a^2+r^2\right)}{3-8 a^2 P \pi }-\gamma  \text{Log}\left( \left(a^2 r^{3 \omega } \left(-4 Q^3 \left(3+4 P \pi
	r^2\right)+r^3 \left(-3+8 P \pi  r^2\right)\right)+3 r \left(r^{4+3 \omega }+\right.\right.\right.}\notag\\ 
&{\left.\left.8 P \pi  r^{6+3 \omega }+3 r^3 \beta  \omega +Q^3 \left(-2 r^{1+3 \omega }+3 \beta  (1+\omega )\right)\right)\right)^2/r^{-2-6 \omega
	}\left(48 \pi  \left(3-8 a^2 P \pi \right)\right.}\notag\\ 
&{\left.\left. \left(a^2+r^2\right) \left(Q^3+r^3\right)^2\right)\right)}.
\end{align}
\begin{figure}[H]
	\centering
	\includegraphics[width=0.5\linewidth, height=5cm]{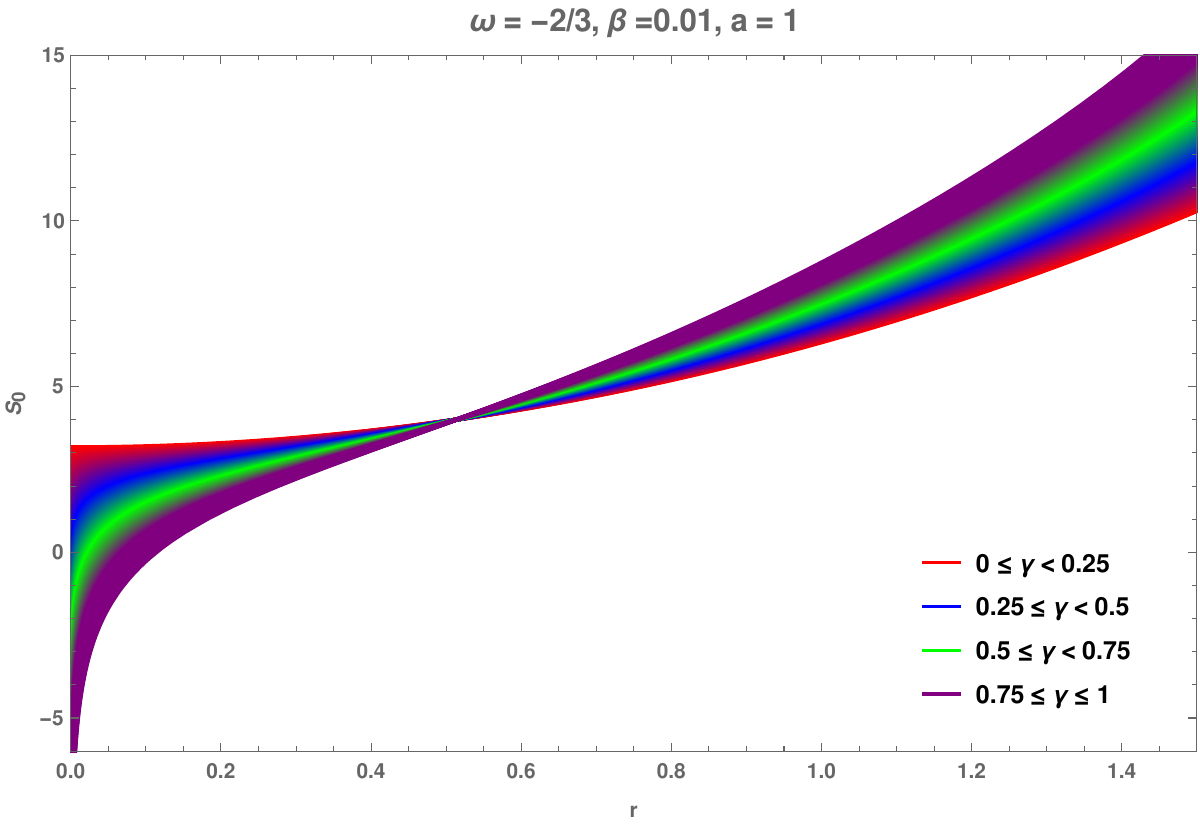}
	\caption{The variation of the corrected entropy with \(r\) for different values of the correction parameter.}
	\label{Fig1}
	\end{figure}
	The graph \ref{Fig1} illustrates how entropy behaves with and without corrections. When the system is at equilibrium (with $ \gamma = 0 $), entropy remains positive and increases steadily. This aligns with the second law of black hole thermodynamics, which states that entropy of a black hole should always increase. However, when $ \gamma $ is not zero, smaller black holes can show negative entropy, and this effect intensifies as $ \gamma $ rises. For larger black holes, corrected entropy continues to be positive, similar to what is observed without corrections. This finding is significant as it highlights the impact of thermal fluctuations on the entropy of smaller black holes.

We examine the impact of a quintessence field on a nonlinear magnetic charged rotating AdS black hole with \(a = 0\). The critical points are identified by applying the critical condition.
\begin{equation}\label{Eq 7}
\begin{aligned}
\left.\frac{\partial T}{\partial r}\right|_{\beta, P, \omega} = 0, \quad \left.\frac{\partial^2 T}{\partial r^2}\right|_{\beta, P, \omega} = 0,
	\end{aligned}
\end{equation}
we derive the following equations, where \(r_c\) and \(P_c\) denote the critical horizon radius and critical pressure, respectively.
\begin{equation}\label{Eq 8}
\begin{aligned}
P_c=& \left(3 Q^6 \beta  \left(2+5 \omega +3 \omega ^2\right)+3 Q^3 \beta  \left(5+7 \omega +6 \omega ^2\right) r_c^3+3 \beta  \omega
	(2+3 \omega ) r_c^6-2 Q^6 r_c^{1+3 \omega }-\right.\\
&\left.10 Q^3 r_c^{4+3 \omega }+r_c^{7+3 \omega }\right)r_c^{-3 (2+\omega )}/\left(8 \pi  \left(4 Q^3+r_c^3\right)\right),
	\end{aligned}
\end{equation}
the critical radius \( r_c \) is determined as the solution to the equation
\begin{equation}\label{Eq.9}
\begin{aligned}
&\left(4 Q^6 \left(9 \beta  \left(4+12 \omega +11 \omega ^2+3 \omega ^3\right)-10 r_c^{1+3 \omega }\right)+r_c^6 \left(9 \beta  \omega
	\left(2+5 \omega +3 \omega ^2\right)+2 r_c^{1+3 \omega }\right)\right.\\
&\left.-Q^3 r_c^3 \left(-9 \beta  \left(10+17 \omega +22 \omega ^2+15 \omega ^3\right)+56 r_c^{1+3 \omega }\right)\right)=0.
	\end{aligned}
\end{equation}
For a nonlinear magnetic charged rotating AdS black hole surrounded by quintessence ($ a \neq 0 $), the critical point can be identified by the conditions \(\left.\frac{\partial T}{\partial S}\right|_{\beta, P, \omega, j} = 0\) and \(\left.\frac{\partial^2 T}{\partial S^2}\right|_{\beta, P, \omega, j} = 0\). This requires expressing the angular momentum parameter \( a \) in terms of the entropy \( S \), pressure \( P \), angular momentum \( j \), magnetic charge \( Q \), quintessence parameter \( \beta \), and the state parameter \( \omega \).
\begin{equation}\label{Eq 10}
\begin{aligned}
{a=\dfrac{6 j \sqrt{\pi } S^{\frac{3}{2}+\frac{3 \omega }{2}}}{3 \pi ^{3/2} Q^3 S^{\frac{1}{2}+\frac{3 \omega }{2}}+8 P \pi ^{3/2}
		Q^3 S^{\frac{3}{2}+\frac{3 \omega }{2}}+3 S^{2+\frac{3 \omega }{2}}+8 P S^{3+\frac{3 \omega }{2}}-3 \pi ^{2+\frac{3 \omega }{2}} Q^3 \beta -3 \pi
		^{\frac{1}{2}+\frac{3 \omega }{2}} S^{3/2} \beta }.}
	\end{aligned}
\end{equation}	
The Hawking temperature is precisely obtained by utilizing the equations presented in Eqs. \ref{Eq 3} and \ref{Eq 10}. 
\begin{equation}\label{Eq 11}
\begin{aligned}{T=\dfrac{S^{-1-\frac{3 \omega }{2}} \left(T_1 T_2+6 j^2 \pi  S^{2+3 \omega } \left(T_3+T_4\right)\right)}{4 \sqrt{\pi } \left(\pi
		^{3/2} Q^3+S^{3/2}\right)^2 T_1},}	
	\end{aligned}
	\end{equation}
 the expressions for \(T_1\), \(T_2\), \(T_3\), and \(T_4\) are provided in the appendix.
 	
Using the earlier critical conditions and with \(\omega = -\frac{2}{3}\), we derive the analytical expression for the critical point. 
	\begin{equation}\label{Eq 12}
\begin{aligned}
&{P_c=\dfrac{\left(P_1+P_2+P_3+P_4\right)}{98304 \pi ^{7/2} Q^6 \left(3 \sqrt{\pi }-32 \beta \right)^2}},  \\ &{S_c=\dfrac{-27 \pi ^{3/2} Q^3 \beta +\left(3 \sqrt{\pi }-32 \beta \right) \sqrt{Z_2}+\left(3 \sqrt{\pi }-32 \beta \right) \sqrt{Z_3}}{6
		\sqrt{\pi }-64 \beta }},\\&{T_c=\frac{T_1 T_2+6 j^2 \pi  \left(T_3+T_4\right)}{4 \sqrt{\pi } \left(\pi ^{3/2} Q^3+S_c^{3/2}\right){}^2 T_1}},
\end{aligned}
\end{equation}
with \(P_1\), \(P_2\), \(P_3\), \(P_4\), \(Z_2\), and \(Z_3\) defined in the appendix, and by applying the same procedure, we derive the analytical expression for the critical point across all values of \(\omega\).

In classical thermodynamics, the physical mass \(m\) of a black hole can be interpreted as the enthalpy \(H\) rather than the total energy of spacetime \cite{21}. Therefore, by utilizing Eqs. \ref{Eq.2} and \ref{Eq 11}, we can derive the Gibbs free energy.
	\begin{equation}\label{Eq.13}
\begin{aligned}
{G= m- T S = \frac{S^{-\frac{3}{2} (1+\omega )} \left(M_1+6 j^2 \pi ^2 S^{2+3 \omega } \left(M_2-M_3\right)\right)}{6 \sqrt{\pi } \left(\pi
		^{3/2} Q^3+S^{3/2}\right)^2 \left(8 P S^{\frac{3 (1+\omega )}{2}}+3 S^{\frac{1}{2}+\frac{3 \omega }{2}}-3 \pi ^{\frac{1}{2}+\frac{3 \omega }{2}}
		\beta \right)^2}-T S},
\end{aligned}
\end{equation}
where \(M_1\), \(M_2\), and \(M_3\) are defined in the appendix.

A standard approach to investigating the critical behavior of a black hole is through the analysis of its heat capacity, which is given by the following expression:
	\begin{equation}\label{Eq 14}
\begin{aligned}
C_{P,j,Q,\beta,\omega} = T \left( \frac{\partial S}{\partial T} \right)_{P,j,Q,\beta,\omega}.
\end{aligned}
\end{equation}
\begin{table}[h!]
	\centering
	\begin{tabular}{|c|c|c|c|c|}
		\hline
	\(\beta\)& \(0.003\) & \(0.03\) & \(0.04\) & \(0.1\) \\ \hline
		\(P_c\) & \(0.001783\) & \(0.001408\) & \( 0.001274\) & \(0.000549\) \\ \hline
		\(S_c\) & \(45.269151\) & \(48.403707\) & \(49.795653 \) & \(63.317609\) \\ \hline
		\(T_c\) & \(0.032265\) & \(0.025129\) & \(0.022467\) & \(0.006116\) \\ \hline
		\(G_c\) & \(0.878647\) & \(0.921519\) & \(0.940062\) & \(1.108196\) \\ \hline
	\end{tabular}
	\caption{Values of \(P_c\), \(S_c\), \(T_c\), and \(G_c\) for different \(\beta\) values with j=0.8, Q=1 and $ \omega$ =-2/3.}\label{1}
\end{table}
The table \ref{1} illustrates the critical thermodynamic properties of a nonlinear magnetic charged rotating AdS black hole  under varying values of the quintessence field parameter \(\beta\). As \(\beta\) increases, the critical pressure and temperature decrease, indicating that the quintessence field lowers the conditions required for phase transitions, effectively softening the black hole system. In contrast, the critical entropy rises with increasing \(\beta\), suggesting an expansion of the black hole’s event horizon at criticality. Additionally, the Gibbs free energy also increases, signifying enhanced thermodynamic stability with higher \(\beta\). Overall, the quintessence field significantly influences the black hole’s phase behavior, resulting in reduced critical pressure and temperature, but greater entropy and stability.

To investigate the influence of the quintessence field on the phase transitions of a nonlinear magnetic charged rotating AdS black hole, we analyze various critical behaviors. For the analysis of \(T - S\), \(G - S\), and \(C - S\) behaviors, we use Eqs. \ref{Eq 11}, \ref{Eq.13}, and \ref{Eq 14}, respectively.
\begin{figure}[H]
	\centering
	\includegraphics[width=0.31\linewidth, height=4cm]{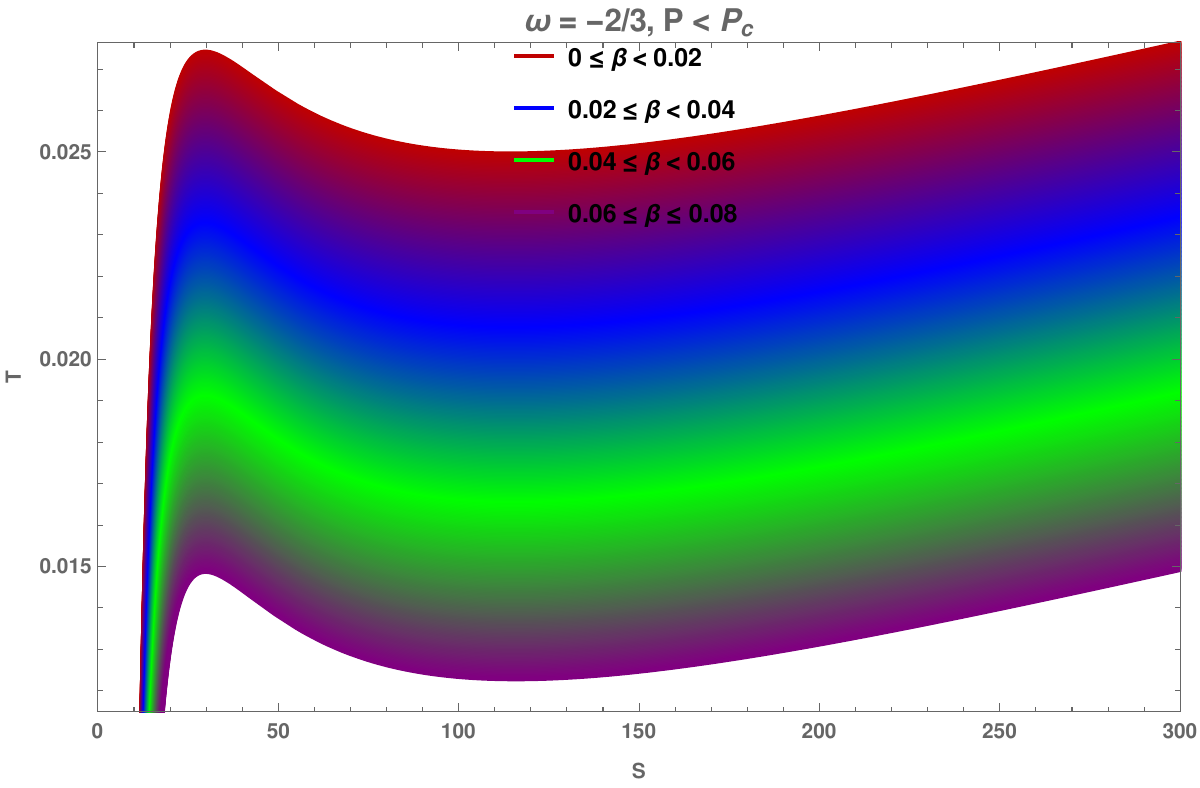}\hspace{0.3cm} \includegraphics[width=0.31\linewidth, height=4cm]{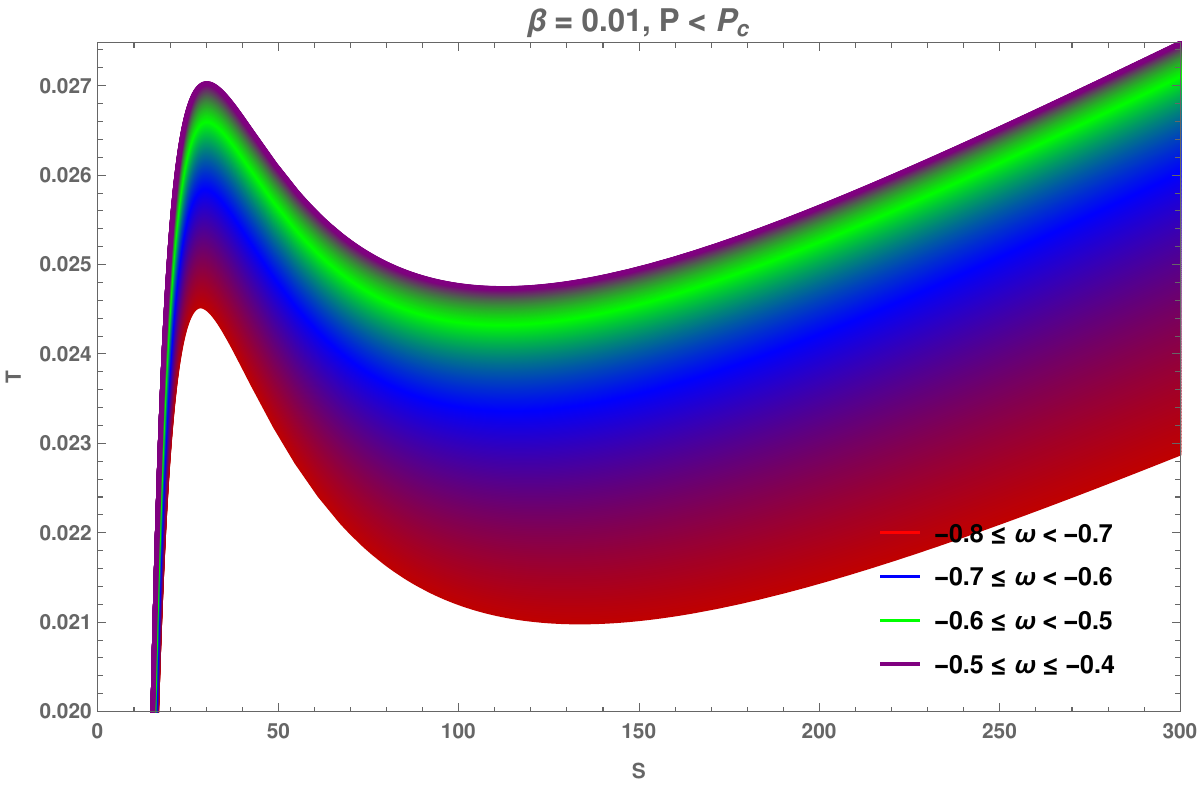}
	\hspace{0.3cm} \includegraphics[width=0.31\linewidth, height=4cm]{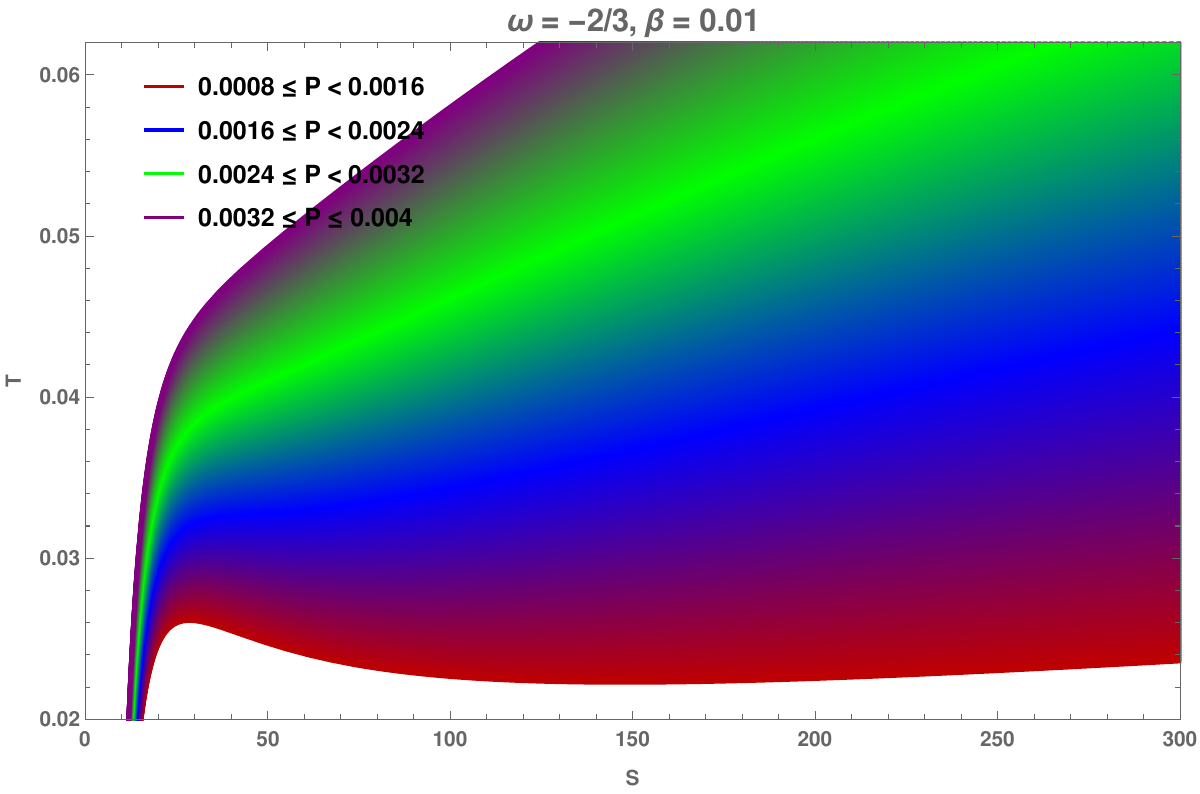}
	\caption{Temperature vs. Entropy for varying $\beta$ and  $\omega$.}
	\label{fig2}
\end{figure}

\begin{figure}[H]
	\centering
	\includegraphics[width=0.31\linewidth, height=4cm]{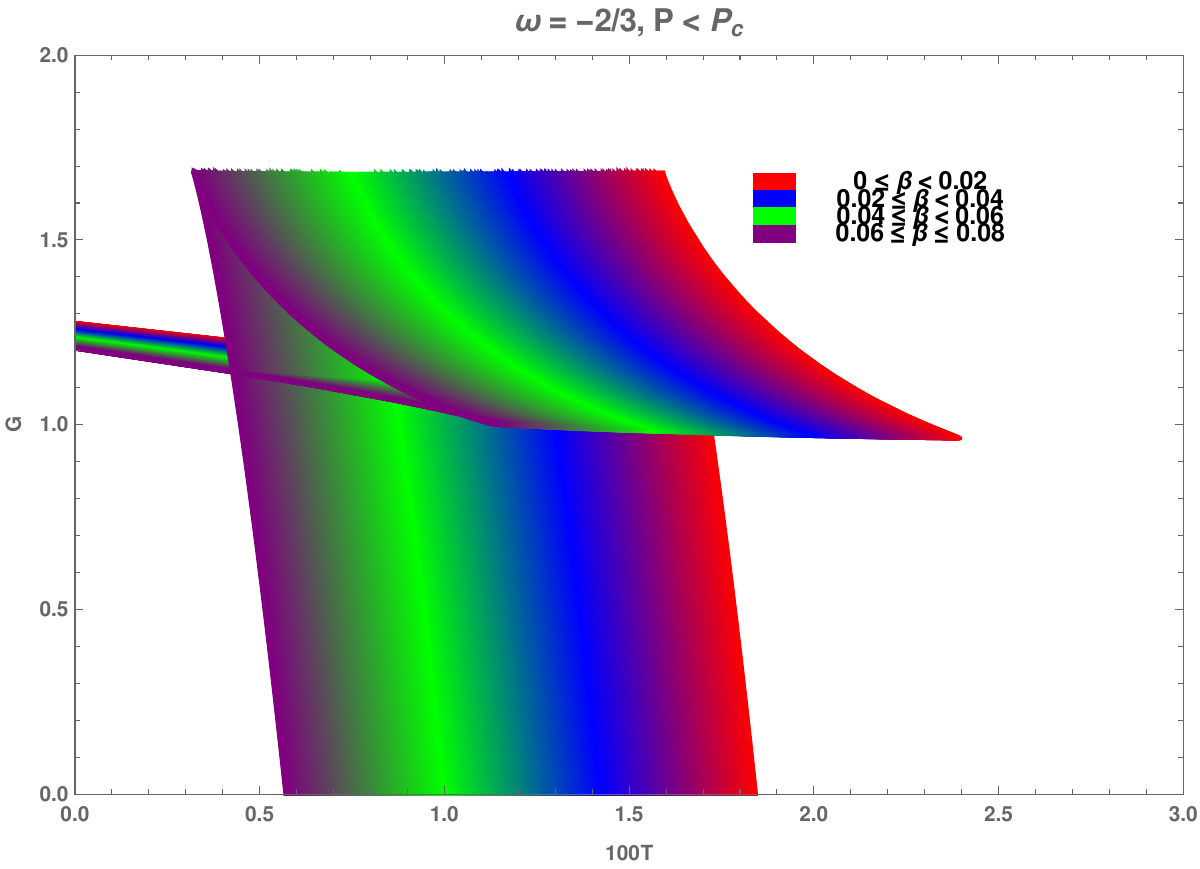}\hspace{0.3cm}
	\includegraphics[width=0.31\linewidth, height=4cm]{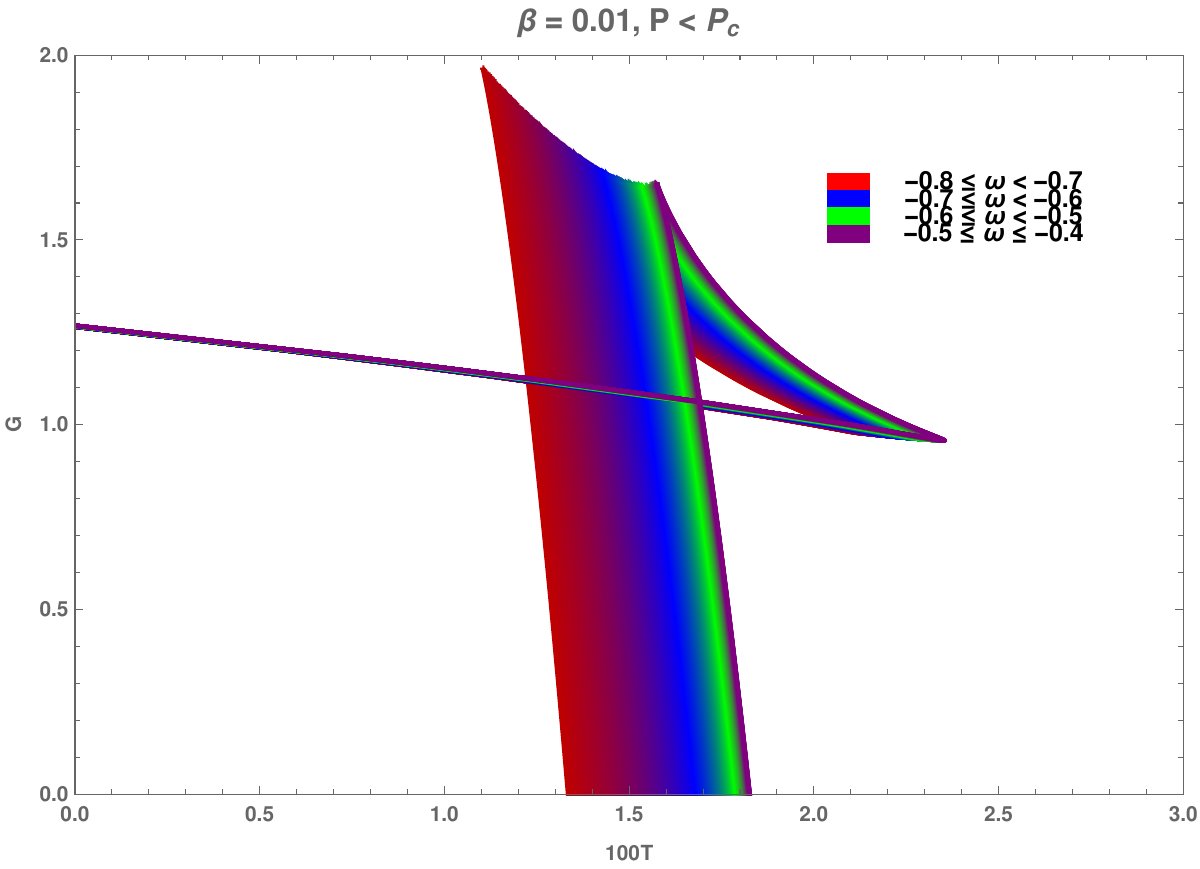}\hspace{0.3cm} \includegraphics[width=0.31\linewidth, height=4cm]{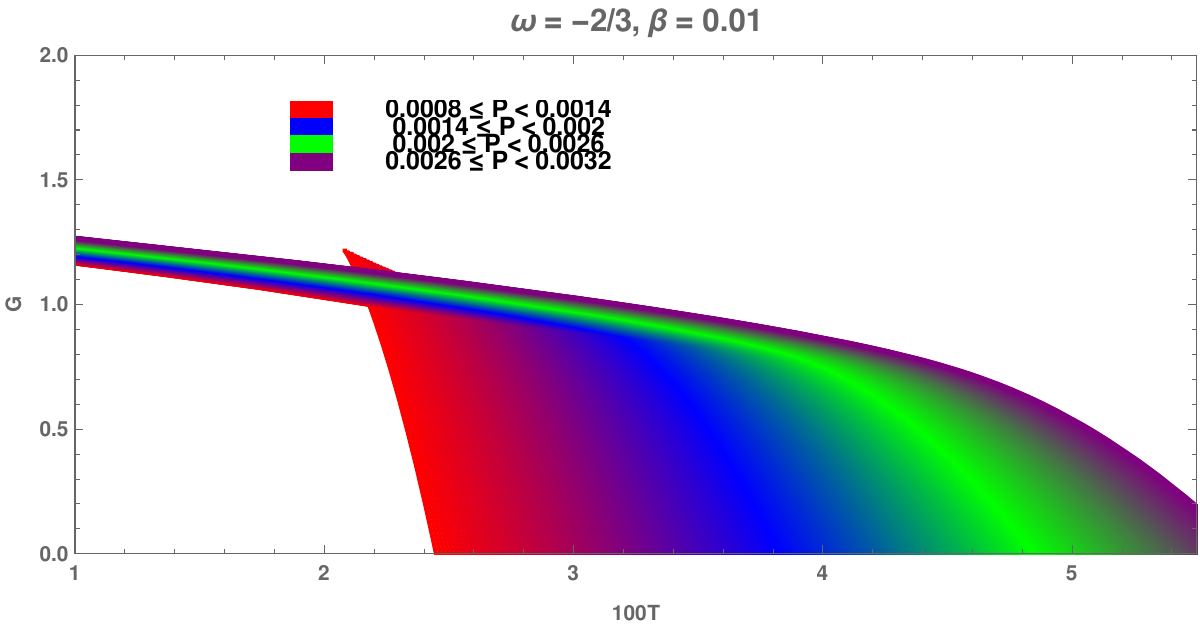}
	\caption{Gibbs Free Energy vs. Temperature for varying  $\beta$ and  $\omega$.}
	\label{fig3}
\end{figure}
\begin{figure}[H]
	\centering
	\includegraphics[width=0.31\linewidth, height=4cm]{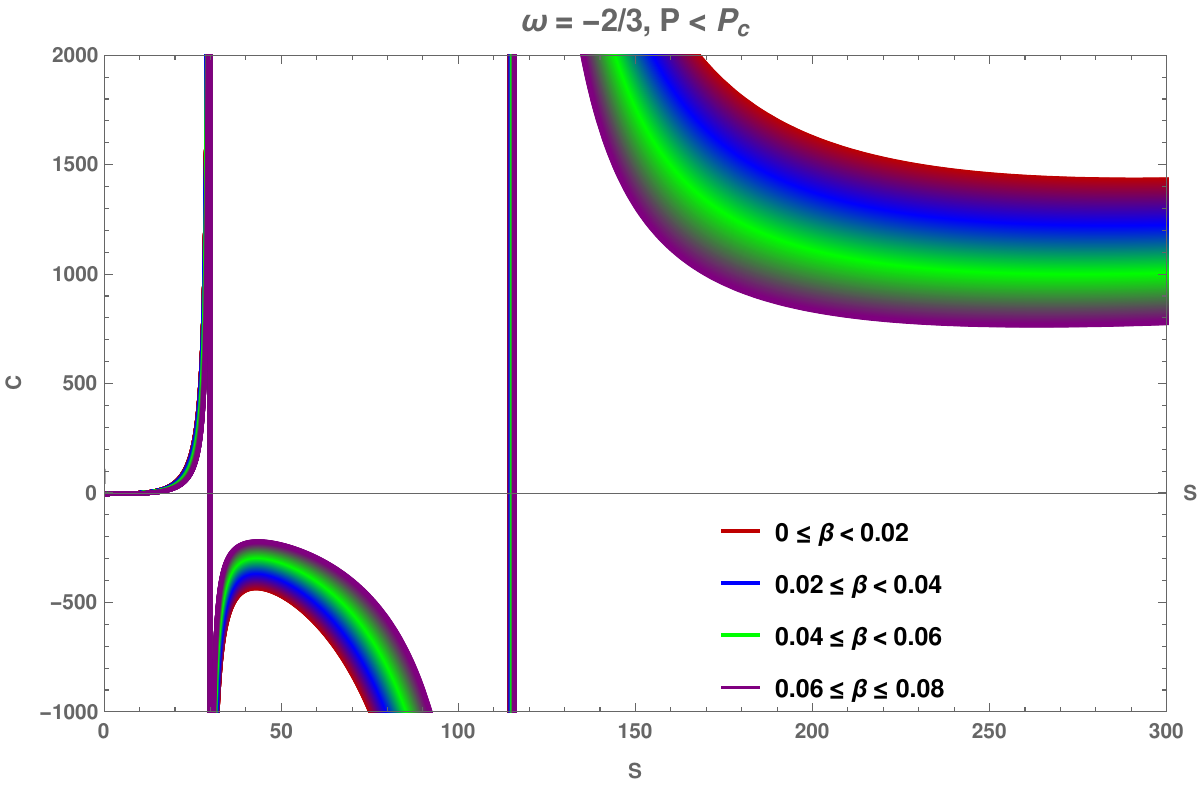}\hspace{0.3cm} \includegraphics[width=0.31\linewidth, height=4cm]{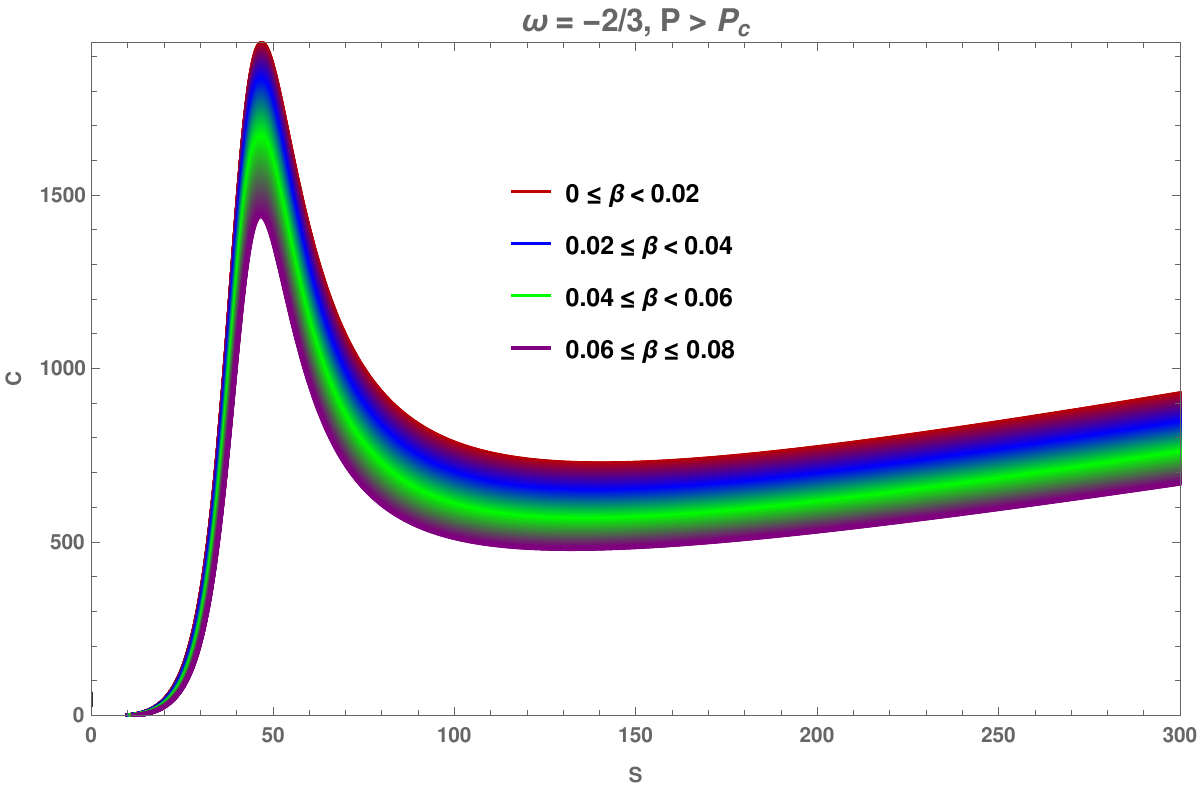}
	\hspace{0.3cm} \includegraphics[width=0.31\linewidth, height=4.1cm]{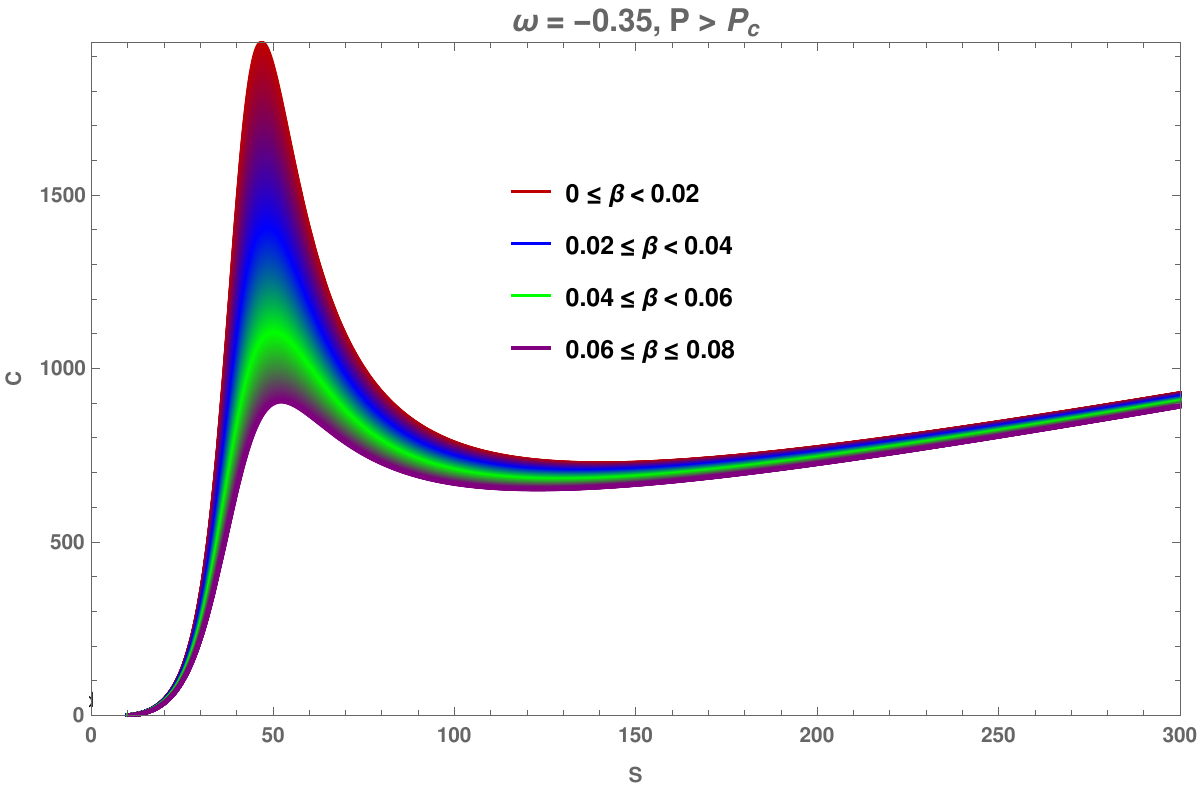}
	\caption{Heat Capacity vs. Entropy for Different Values of \(\beta\) and \(\omega\).}
	\label{fig4}
\end{figure}

The plots illustrate that \(T - S\) criticality is present for \(P < P_c\) . In Fig. \ref{fig2}, it is clearly shown that with a fixed parameter \(\omega\), an increase in the quintessence parameter \(\beta\) results in a decrease in the local maximum and minimum points of the temperature \(T\). This indicates that as \(\beta\) grows, the extremes of \(T\) are suppressed. On the other hand, when \(\omega\) is increased while keeping \(\beta\) constant, there is a noticeable shift in these local extrema, with the maximum and minimum points moving toward higher values of \(T\). This suggests that \(\omega\) has a significant impact on the critical temperature behavior. For values of \(P\) greater than \(P_c\), the oscillatory behavior in \(T - S\) plots disappears. Additionally, when \(P\) is less than \(P_c\), the Gibbs free energy exhibits a pronounced swallowtail shape, as depicted in Fig. \ref{fig3}. This swallowtail pattern highlights the critical points of the system's thermodynamic phase transitions. Notably, a reduction in the parameter \(\beta\) causes the intersection point on the Gibbs free energy plot to shift toward higher values of  \(G\) and  \(T\). However, when \(P\) exceeds \(P_c\), the distinctive swallowtail pattern in the Gibbs free energy curve completely vanishes. Furthermore, it becomes apparent that increasing \(\omega\) leads to a higher temperature at the intersection point, while the minimum temperature shows a decreasing trend, as illustrated in Fig.  \ref{fig3}. Moreover, the heat capacity \(C\) exhibits discontinuities at two distinct points when \(P < P_c\) (see Fig. \ref{fig4}), signaling thermodynamic instability within the region bounded by these divergence points. This instability manifests as abrupt changes in the heat capacity, highlighting critical transitions in the system. Outside this unstable region, the system regains thermodynamic stability. When \(P\) surpasses \(P_c\), the discontinuities in \(C\) disappear, and the previously observed phase transition ceases to occur. All the plots  demonstrate that the nonlinear magnetic charged
rotating AdS black hole is influenced by both the \(\beta\) and \(\omega\) parameters, showing complex dependencies that affect the system's overall thermodynamic behavior.

This study focuses on exploring how thermal fluctuations affect a nonlinear magnetic charged
rotating AdS black hole surrounded by quintessence. The findings reveal that thermal fluctuations significantly intensify, especially for smaller black holes. Additionally, this effect becomes more pronounced as the corrected parameter $\gamma$ increases across various values of the quintessence parameter $\beta$. Through critical conditions and detailed calculations, we identify the analytical expression for the critical points of these types of AdS black holes. We demonstrate that for certain values of $\omega$ and $\beta$ with $P < P_c$, the Gibbs free energy-temperature (G-T) and temperature-entropy (T-S) diagrams exhibit swallowtail behavior and oscillatory parts, respectively, indicating a first-order phase transition known as the small–large black hole phase transition. However, when $P > P_c$, this phase transition does not occur.
The analysis concludes that the presence of a quintessence field around a nonlinear magnetic charged
rotating AdS black hole does not affect the occurrence of the small–large black hole phase transition, suggesting that these black holes undergo a phase transition similar to that of a van der Waals fluid. A key finding of this study is the sensitivity to the parameters $\omega$ and $\beta$. As $\beta$ increases, it significantly impacts the temperature, leading to a substantial shift toward lower values for the two extreme points, while the intersection point of the Gibbs free energy moves to lower temperatures and lower Gibbs free energy values. Conversely, increasing $\omega$ reduces both the phase transition point and the coexistence region of the two phases, shifting the temperature to higher values and the intersection point to higher temperatures and Gibbs free energy.

\label{7}

\appendix

\textbf{\Large{Appendix}}\

\noindent\({A_1=- \left(2460375 j^6 \pi ^3+8748000 j^4 \pi ^2 Q^3 \left(3 \sqrt{\pi }-8 \beta \right)+3240 j^2 \pi  Q^6 \left(27360 \pi -122880
	\sqrt{\pi } \beta +40960\beta ^2\right.\right.}\\
{\left.\left. +2187 \pi ^{3/2} Q^3 \beta ^2\right)+64 Q^9 \left(1506816 \pi ^{3/2}-8819712 \pi  \beta +5603328 \sqrt{\pi } \beta ^2+295245
	\pi ^2 Q^3 \beta ^2-262144 \beta ^3\right)\right)}\\
{16 \pi ^{9/2}/\left(3 \sqrt{\pi }-32 \beta \right)^3}\),

\noindent\({A_2=- 256\left(18225 j^4 \pi ^5+90720 j^2 \pi ^{9/2} Q^3+112896 \pi ^4 Q^6+69120 j^2 \pi ^4 Q^3 \beta +172032 \pi ^{7/2} Q^6 \beta
	+\right.}\\
{\left.65536 \pi ^3 Q^6 \beta ^2\right)^3/\left(3 \sqrt{\pi }-32 \beta \right)^6+A_1^2}\),

\noindent\({A_3=\left(A_1 \left(9216 \pi ^{5/2} Q^3-104448 \pi ^2 Q^3 \beta +65536 \pi ^{3/2} Q^3 \beta ^2+4374 \pi ^3 Q^6 \beta ^2+1080 j^2
	\left(3 \pi ^3-32 \pi ^{5/2} \beta \right)\right.\right.}\\
{\left.+9 2^{2/3} \pi  \left(A_1+\sqrt{A_2}\right){}^{1/3}-192\ 2^{2/3} \sqrt{\pi } \beta  \left(A_1+\sqrt{A_2}\right){}^{1/3}+1024\ 2^{2/3} \beta
	^2 \left(A_1+\sqrt{A_2}\right){}^{1/3}\right)+8\ 2^{1/3} \pi ^3 }\\
{\left(135 j^2 \pi +16 Q^3 \left(21 \sqrt{\pi }+16 \beta \right)\right)^2 \left(A_1+\sqrt{A_2}\right){}^{2/3}+\left(9216 \pi ^{5/2} Q^3-104448
	\pi ^2 Q^3 \beta +65536 \pi ^{3/2} Q^3 \beta ^2\right.}\\
{+4374 \pi ^3 Q^6 \beta ^2+1080 j^2 \left(3 \pi ^3-32 \pi ^{5/2} \beta \right)+9\ 2^{2/3} \pi  \left(A_1+\sqrt{A_2}\right){}^{1/3}-192\ 2^{2/3}
	\sqrt{\pi } \beta  \left(A_1+\sqrt{A_2}\right){}^{1/3}}\\
{\left.\left.+1024\ 2^{2/3} \beta ^2 \left(A_1+\sqrt{A_2}\right){}^{1/3}\right) \sqrt{A_2}\right)/\left(\left(3 \sqrt{\pi }-32 \beta \right)^2
	\left(A_1+\sqrt{A_2}\right)\right)}\),

\noindent\({A_4=\left(-324 \sqrt{6} \pi ^3 Q^3 \beta  \left(Q^3 \left(2304 \pi -26112 \sqrt{\pi } \beta +16384 \beta ^2+729 \pi ^{3/2} Q^3 \beta
	^2\right)+270 j^2 \right.\right.}\\
{\left.\left(3 \pi ^{3/2}-32 \pi  \beta \right)\right) \left(A_1+\sqrt{A_2}\right){}^{1/3}-\left(3 \sqrt{\pi }-32 \beta \right) \left(145800\
	2^{1/3} j^4 \pi ^5+903168\ 2^{1/3} \pi ^4 Q^6\right.}\\
{+1376256\ 2^{1/3} \pi ^{7/2} Q^6 \beta +4 \pi ^3 Q^6 \beta ^2 \left(131072\ 2^{1/3}-2187 \left(A_1+\sqrt{A_2}\right){}^{1/3}\right)-2160 j^2
}\\
{\left(-336 2^{1/3} \pi ^{9/2} Q^3-256\ 2^{1/3} \pi ^4 Q^3 \beta +3 \pi ^3 \left(A_1+\sqrt{A_2}\right){}^{1/3}-32 \pi ^{5/2} \beta  \left(A_1+\sqrt{A_2}\right){}^{1/3}\right)-}\\
{18432 \pi ^{5/2} Q^3 \left(A_1+\sqrt{A_2}\right){}^{1/3}+208896 \pi ^2 Q^3 \beta  \left(A_1+\sqrt{A_2}\right){}^{1/3}-131072 \pi ^{3/2} Q^3
	\beta ^2 \left(A_1+\sqrt{A_2}\right){}^{1/3}}\\
{\left.\left.+9\ 2^{2/3} \pi  \left(A_1+\sqrt{A_2}\right){}^{2/3}-192\ 2^{2/3} \sqrt{\pi } \beta  \left(A_1+\sqrt{A_2}\right){}^{2/3}+1024\ 2^{2/3}
	\beta ^2 \left(A_1+\sqrt{A_2}\right){}^{2/3}\right) \sqrt{A_3}\right)}\\
{\left/\left(\left(3 \sqrt{\pi }-32 \beta \right)^3 \left(A_1+\sqrt{A_2}\right){}^{1/3} \sqrt{A_3}\right)\right.}\),

\noindent\({T_1=\left(\pi ^{3/2} Q^3 S^{\frac{1}{2}+\frac{3 \omega }{2}} (3+8 P S)+S^{2+\frac{3 \omega }{2}} (3+8 P S)-3 \pi ^{2+\frac{3 \omega
		}{2}} Q^3 \beta -3 \pi ^{\frac{1}{2}+\frac{3 \omega }{2}} S^{3/2} \beta \right)^2}\),

\noindent\({T_2=\left(-2 \pi ^3 Q^6 S^{\frac{1}{2}+\frac{3 \omega }{2}}+\pi ^{3/2} Q^3 S^{2+\frac{3 \omega }{2}} (-1+8 P S)+S^{\frac{7}{2}+\frac{3
			\omega }{2}} (1+8 P S)+3 \pi ^{\frac{1}{2}+\frac{3 \omega }{2}} S^3 \beta  \omega +3 \pi ^{\frac{7}{2}+\frac{3 \omega }{2}} Q^6 \right.}\\
{\left.\beta  (1+\omega )+3 \pi ^{2+\frac{3 \omega }{2}} Q^3 S^{3/2} \beta  (1+2 \omega )\right)}\),

\noindent\({T_3=-6 \pi ^4 Q^6 S^{\frac{1}{2}+\frac{3 \omega }{2}} (3+8 P S)-\pi  S^{\frac{7}{2}+\frac{3 \omega }{2}} (3+8 P S)^2-2 \pi ^{5/2}
	Q^3 S^{2+\frac{3 \omega }{2}} \left(27+120 P S+128 P^2 S^2\right)}\),

\noindent\({T_4=3 \pi ^{\frac{3 (1+\omega )}{2}} S^3 \beta  \omega  (9 \omega +8 P S (2+3 \omega ))+3 \pi ^{\frac{3 (3+\omega )}{2}} Q^6 \beta
	(1+\omega ) (9 \omega +8 P S (2+3 \omega ))+3 \pi ^{3+\frac{3 \omega }{2}} Q^3}\\
{ S^{3/2} \beta  \left(9 \left(1+\omega +2 \omega ^2\right)+8 P S \left(5+7 \omega +6 \omega ^2\right)\right)}\),



\noindent\({P_1=-729 \pi ^{7/2} Q^3 \beta  \left(120 j^2 \left(3 \pi -32 \sqrt{\pi } \beta \right)+Q^3 \left(960 \sqrt{\pi }-10240 \beta +243
	\pi  Q^3 \beta ^2\right)\right)-3 \left(3 \sqrt{\pi }-32 \beta \right)^3}\\
{ Z_2^{3/2}-9 \left(3 \sqrt{\pi }-32 \beta \right)^2 Z_2 \left(9 \pi ^{3/2} Q^3 \beta +\left(3 \sqrt{\pi }-32 \beta \right) \sqrt{Z_3}\right)}\),

\noindent\({P_2= \left(3240 j^2 \left(3 \pi ^{3/2}-32 \pi  \beta \right)+Q^3 \left(28224 \pi -325632 \sqrt{\pi } \beta +262144 \beta ^2+10935
	\pi ^{3/2} Q^3 \beta ^2\right)\right) \sqrt{Z_3}}\\
{\pi ^{3/2} \left(3 \sqrt{\pi }-32 \beta \right)}\),

\noindent\({P_3=-81 \pi ^{3/2} Q^3 \left(3 \sqrt{\pi }-32 \beta \right)^2 \beta  Z_3-3 \left(3 \sqrt{\pi }-32 \beta \right)^3 Z_3^{3/2}}\),

\noindent\({P_4=\left(3 \sqrt{\pi }-32 \beta \right) \sqrt{Z_2} \left(\pi ^{3/2} \left(3240 j^2 \left(3 \pi ^{3/2}-32 \pi  \beta \right)+Q^3
	\left(28224 \pi -325632 \sqrt{\pi } \beta +262144 \beta ^2\right.\right.\right.}\\
{\left.\left.\left.+10935 \pi ^{3/2} Q^3 \beta ^2\right)\right)-162 \pi ^{3/2} Q^3 \left(3 \sqrt{\pi }-32 \beta \right) \beta
	\sqrt{Z_3}-9 \left(3 \sqrt{\pi }-32 \beta \right)^2 Z_3\right)}\),

\noindent\({Z_1=16 \pi ^{9/2} \left(-\left(2460375 j^6 \pi ^3+8748000 j^4 \pi ^2 Q^3 \left(3 \sqrt{\pi }-8 \beta \right)+3240 j^2 \pi  Q^6 \left(27360
	\pi -122880 \sqrt{\pi } \beta \right.\right.\right.}\\
{\left.+40960 \beta ^2+2187 \pi ^{3/2} Q^3 \beta ^2\right)+64 Q^9 \left(1506816 \pi ^{3/2}-8819712 \pi  \beta +5603328 \sqrt{\pi } \beta ^2+295245
	\pi ^2 Q^3 \beta ^2\right.}\\
{\left.\left.-262144 \beta ^3\right)\right)/\left(\left(3 \sqrt{\pi }-32 \beta \right)^3\right)+\surd \left(\left(-\left(135 j^2 \pi +16 Q^3
	\left(21 \sqrt{\pi }+16 \beta \right)\right)^6+\left(2460375 j^6 \pi ^3+8748000\right.\right.\right.}\\
{ j^4 \pi ^2 Q^3 \left(3 \sqrt{\pi }-8 \beta \right)+3240 j^2 \pi  Q^6 \left(27360 \pi -122880 \sqrt{\pi } \beta +40960 \beta ^2+2187 \pi ^{3/2}
	Q^3 \beta ^2\right)+64 Q^9}\\
{\left.\left.\left.\left. \left(1506816 \pi ^{3/2}-8819712 \pi  \beta +5603328 \sqrt{\pi } \beta ^2+295245 \pi ^2 Q^3 \beta ^2-262144 \beta ^3\right)\right)^2\right)/\left(\left(3
	\sqrt{\pi }-32 \beta \right)^6\right)\right)\right)}\),

\noindent\({Z_2=\left(8\ 2^{1/3} \pi ^3 \left(135 j^2 \pi +16 Q^3 \left(21 \sqrt{\pi }+16 \beta \right)\right)^2+2 \pi ^{3/2} \left(540 j^2 \left(3
	\pi ^{3/2}-32 \pi  \beta \right)+Q^3 \left(4608 \pi -52224  \right.\right.\right.}\\
{\left.\left.\left.\sqrt{\pi } \beta+32768 \beta ^2+2187 \pi ^{3/2} Q^3 \beta ^2\right)\right) Z_1^{1/3}+2^{2/3} \left(3 \sqrt{\pi }-32 \beta \right)^2 Z_1^{2/3}\right)/\left(6
	\left(3 \sqrt{\pi }-32 \beta \right)^2 Z_1^{1/3}\right)}\),

\noindent\({Z_3=\left(-4 \pi ^{3/2} Z_1^{1/3} \left(81 \pi ^{3/2} Q^3 \beta  \left(270 j^2 \left(3 \pi ^{3/2}-32 \pi  \beta \right)+Q^3 \left(2304
	\pi -26112 \sqrt{\pi } \beta +16384 \beta ^2+729  \right.\right.\right.\right.}\\
{\left.\left.\pi ^{3/2}Q^3 \beta ^2\right)\right)-\left(3 \sqrt{\pi }-32 \beta \right) \left(540 j^2 \left(3 \pi ^{3/2}-32 \pi  \beta \right)+Q^3 \left(4608
	\pi -52224 \sqrt{\pi } \beta +32768 \beta ^2+2187 \pi ^{3/2} \right.\right.}\\
{\left.\left.\left.Q^3 \beta ^2\right)\right) \sqrt{Z_2}\right)-8\ 2^{1/3} \pi ^3 \left(3 \sqrt{\pi }-32 \beta \right) \left(135 j^2 \pi +16
	Q^3 \left(21 \sqrt{\pi }+16 \beta \right)\right)^2 \sqrt{Z_2}-2^{2/3} \left(3 \sqrt{\pi }-32 \beta \right)^3}\\
{\left. Z_1^{2/3} \sqrt{Z_2}\right)/\left(6 \left(3 \sqrt{\pi }-32 \beta \right)^3 Z_1^{1/3} \sqrt{Z_2}\right)}\),

\noindent\({M_1=\left(\pi ^{3/2} Q^3+S^{3/2}\right)^3 \left(8 P S^{\frac{3 (1+\omega )}{2}}+3 S^{\frac{1}{2}+\frac{3 \omega }{2}}-3 \pi ^{\frac{1}{2}+\frac{3
			\omega }{2}} \beta \right)^3}\),

\noindent\({M_2=4 \pi ^{3/2} Q^3 S^{\frac{1}{2}+\frac{3 \omega }{2}} (3+8 P S)^2+S^{2+\frac{3 \omega }{2}} (3+8 P S)^2-3 \pi ^{\frac{1}{2}+\frac{3
			\omega }{2}} S^{3/2} \beta  (9 \omega +8 P S (4+3 \omega ))}\),

\noindent\({M_3=3 \pi ^{2+\frac{3 \omega }{2}} Q^3 \beta  (9 (1+\omega )+8 P S (7+3 \omega ))}\).


\begin{thebibliography}{50}
	\footnotesize
\bibitem{1} Bardeen, J. M. (1968). Non-singular general-relativistic gravitational collapse. Proceedings of the International Conference GR5, Tbilisi, USSR.

\bibitem{2} Beato, E., \& Garcia, A. (2000). The Bardeen model as a nonlinear magnetic monopole. Physics Letters B, 493(1-2), 149-152.

\bibitem{3} Bronnikov, K. A., Fabris, J. C., \& Zerbini, S. (2001). Regular black holes and topology change. General Relativity and Gravitation, 33, 2549-2558.

\bibitem{4} Novello, M., Bergliaffa, S. E. P., \& Salim, J. M. (2000). Nonlinear electrodynamics and the acoustic black hole. Physical Review D, 61(4), 045001.

\bibitem{5} Breton, N. (2003). Geodesic structure of the Born–Infeld black hole. Classical and Quantum Gravity, 19(23), 601-612.

\bibitem{6} Kiselev, V. V. (2003). The Black Hole in a Universe with Quintessence. Classical and Quantum Gravity, 20, 1187-1197.

\bibitem{7} Chen, S., Wang, B., \& Su, R. (2008). Critical behavior of charged AdS black holes in the extended phase space. Physical Review D, 77, 124011.

\bibitem{8} Ahmed Daassou,  Rachid Benbrik et Hayat Laassiri (2023), Effect of a cloud of strings and quintessence on a phase transition of charged rotating AdS black holes. Theoretical and Mathematical Physics, 215, 893–908.

\bibitem{9} Laassiri, H., Daassou, A., \& Benbrik, R. (2023). Responses of Small and Large AdS Black Holes to the Collective Influence of Quintessence and String Cloud. Submitted, arXiv:2312.08868.

\bibitem{10} Saleh, M., Thomas, B. B., \& Kofane, T. C. (2018). Thermodynamics of AdS Black Holes. European Physical Journal C, 78, 325.

\bibitem{11} Laassiri, H., Daassou, A., \& Benbrik, R. (2024). Analytical Critical Phenomena of Rotating Regular AdS Black Holes with Dark Energy. International Journal of Modern Physics A 10.1142.

\bibitem{12} Rodrigues, M. E., Silva, M. V. de S., \& Vieira, H. A. (2022). Thermodynamics and Critical Phenomena of Black Holes. Physical Review D, 105, 8.

\bibitem{13} Witten, E. (1998). Anti-de Sitter space and holography. Advances in Theoretical and Mathematical Physics, 2, 253–291. arXiv:hep-th/9802150.

\bibitem{14} Cvetic, M., \& Gubser, S. S. (1999). Phases of R-charged black holes, spinning branes and strongly coupled gauge theories. Journal of High Energy Physics, 04, 024.

\bibitem{15} Maldacena, J. (1999). The large-N limit of superconformal field theories and supergravity. International Journal of Theoretical Physics, 38, 1113–1133. arXiv:hep-th/9711200.

\bibitem{16} Banerjee, R., \& Roychowdhury, D. (2011). Thermodynamics of phase transition in higher dimensional AdS black holes. Journal of High Energy Physics, 11, 004.

\bibitem{17} Niu, C., Tian, Y., \& Wu, X.-N. (2012). Critical phenomena and thermodynamic geometry of Reissner–Nordström–anti-de Sitter black holes. Physical Review D, 85, 024017.

\bibitem{18} Laassiri, H., Daassou, A., \& Benbrik, R. (2024). Thermodynamic Features of AdS Black Holes within the Rastall Gravity and Perfect Fluid Matter Framework. Physica Scripta, 99(7), 5024.

\bibitem{19} Benavides-Gallego, C. A., Abdujabbarov, A., \& Bambi, C. (2020). Rotating and nonlinear magnetic-charged black hole surrounded by quintessence. Physical Review D, 101(4), 044038.

\bibitem{20} Nam, C. H. (2018). Thermodynamics and Phase Transitions of Black Holes. General Relativity and Gravitation, 50, 57.

\bibitem{21} Kastor, D., Ray, S., \& Traschen, J. (2009). Enthalpy and the mechanics of AdS black holes. Classical and Quantum Gravity, 26, 195011.


\end{thebibliography}
\end{document}